\def\year{2018}\relax
\documentclass[letterpaper]{article} 
\usepackage{aaai18}  
\usepackage{times}  
\usepackage{helvet}  
\usepackage{courier}  
\usepackage{url}  
\usepackage[pdftex]{graphicx}   
\usepackage{amsmath}            
\usepackage{accents}            
\usepackage{subfigure}          
\usepackage{booktabs}           
\usepackage{multirow}           
\usepackage{amssymb}            
\usepackage{CJK}                

\begin{document}

\title{A Hierarchical Contextual Attention-based GRU Network \\ for Sequential Recommendation}
\author{Qiang Cui$^1$  \hspace{7mm} Shu Wu$^1$  \hspace{7mm} Yan Huang$^1$ \hspace{7mm} Liang Wang$^{1,2}$ \\
$^1$Center for Research on Intelligent Perception and Computing\\
National Laboratory of Pattern Recognition\\
$^2$Center for Excellence in Brain Science and Intelligence Technology\\
Institute of Automation, Chinese Academy of Sciences\\
{\tt\small cuiqiang2013@ia.ac.cn, \{shu.wu, yhuang, wangliang\}@nlpr.ia.ac.cn}}
\maketitle

\begin{abstract}
    \noindent
    Sequential recommendation is one of fundamental tasks for Web applications.
    Previous methods are mostly based on Markov chains with a strong Markov assumption. Recently, recurrent neural networks (RNNs) are getting more and more popular and has demonstrated its effectiveness in many tasks.
    The last hidden state is usually applied as the sequence's representation to make recommendation.
    Benefit from the natural characteristics of RNN, the hidden state is a combination of long-term dependency and short-term interest to some degrees.
    However, the monotonic temporal dependency of RNN impairs the user's short-term interest. Consequently, the hidden state is not sufficient to reflect the user's final interest.
    In this work, to deal with this problem, we propose a \textbf{H}ierarchical \textbf{C}ontextual \textbf{A}ttention-based \textbf{GRU} (\textbf{HCA}-\textbf{GRU}) network.
    The first level of HCA-GRU is conducted on the input. We construct a contextual input by using several recent inputs based on the attention mechanism. This can model the complicated correlations among recent items and strengthen the hidden state.
    The second level is executed on the hidden state. We fuse the current hidden state and a contextual hidden state built by the attention mechanism, which leads to a more suitable user's overall interest.
    Experiments on two real-world datasets show that HCA-GRU can effectively generate the personalized ranking list and achieve significant improvement.
\end{abstract}

\section{1 Introduction}
    \noindent
    Recently, with the development of Internet, some applications of sequential scene have become numerous and multilateral, such as ad click prediction, purchase recommendation and web page recommendation. A user's behaviors in such applications is a sequence in chronological order, and his following behaviors can be predicted by sequential recommendation methods.
    Taking online shopping for instance, after a user buys an item, the application would predict a list of items that the user might purchase in the near future.

    Traditional sequential methods usually focus on Markov chains \cite{rendle2010factorizing,chen2015personalized,he2016fusing}. However, the underlying strong Markov assumption has difficulty in constructing more effective relationship among various factors.
    Lately, with the great success of deep learning, recurrent neural network (RNN) methods become more and more popular. They have achieved promising performance on different tasks, for example next basket recommendation \cite{yu2016dynamic}, location prediction \cite{liu2016predicting}, product recommendations \cite{cui2016visual,hidasi2016parallel}, and multi-behavioral prediction \cite{liu2017multi}.
    These RNN methods mostly employ a common strategy: using the last hidden state of RNN as the user's final representation then making recommendations to users.

    In recommender systems, a better recommendation should capture both the long-term user-taste and short-term sequential effect \cite{rendle2010factorizing,wang2015learning}.
    They are referred as long-term dependency and short-term interest in this work.
    The hidden state of RNN has both characteristics in nature and is very suitable for recommendation. With the help of gated activation function like long-short term memory \cite{hochreiter1997long} or gated recurrent unit \cite{cho2014learning}, RNN can better capture the long-term dependency. This advantage allows the hidden state to be able to connect previous information for a long time.
    Due to the recurrent structure and fixed transition matrices, RNN holds an assumption that temporal dependency has a monotonic change with the input time steps \cite{liu2017multi}. It assumes that the current item or hidden state is more significant than the previous one. This monotonic assumption is reasonable on the whole sequence, and it also enables hidden state to capture the user's short-term interest to some degrees.


    There is a problem that the monotonic assumption of RNN restricts the modeling of user's short-term interest. Because of this problem, the forementioned common strategy is not sufficient for better recommendation.
    First, short-term interest should discover correlations among items within a local range. This will make the recommendation very responsive for the coming behaviors based on recent behaviors \cite{wang2015learning}.
    Next, for short-term interest in recommender systems, we can not say that one item or hidden state is must more significant than the previous one \cite{liu2017multi}. The correlations are more complicated, but the monotonic assumption can not well distinguish importances of the several recent factors.
    For example, when a user buys clothes online, he might buy it for someone else. Small weights should be provided for such kind of behaviors. Take another example, we can consider the regular three meals a day as a sequence. The connection may be closer among breakfasts instead of between breakfast and lunch or dinner.
    Therefore, the short-term interest should be carefully examined and needs to be integrated with the long-term dependency.


    In this paper, we propose a \textbf{H}ierarchical \textbf{C}ontextual \textbf{A}ttention-based \textbf{GRU} (\textbf{HCA}-\textbf{GRU}) network to address the above problem.
    It can greatly strengthen the user's short-term interest represented in each hidden state, and make a non-linear combination of long-term dependency and short-term interest to obtain user's overall interest.
    The first level of our HCA-GRU is conducted on the input to make hidden states stronger.
    When we input the current item to GRU, we also input a contextual input containing information of previous inputs.
    Inspired by a learning framework for word vectors in Natural Language Processing (NLP) \cite{le2014distributed}, we firstly collect several recent inputs as the contextual information at each time step. Then the attention mechanism is accordingly introduced to assign appropriate weights to select important inputs. The contextual input is a weighted sum of this context.
    The second level of our HCA-GRU is executed on the hidden state to make a fusion of recent hidden states. Each time, the user's overall interest is a combination of the current hidden state and a contextual hidden state.
    Encouraged by the global context memory initialized by all hidden states with average pooling in a Computer Vision (CV) work \cite{liu2017global}, we use several recent hidden states to construct local contextual information. It is used to establish the current contextual hidden state via attention.
    Finally, parameters of HCA-GRU are learned by the Bayesian Personalized Ranking (BPR) framework \cite{rendle2009bpr} and the Back Propagation Through Time (BPTT) algorithm \cite{werbos1990backpropagation}.
    The main contributions are listed as follows:
    \begin{itemize}
        \item
            We propose a novel hierarchical GRU network for sequential recommendation which can combine the long-term dependency and user's short-term interest.
        \item
            We propose to employ contextual attention-based modeling to deal with the monotonic assumption of RNN methods. This can automatically focus on critical information, which greatly strengthens the user's short-term interest.
        \item
            Experiments on two large real-world datasets reveal that the HCA-GRU network is very effective and outperforms the state-of-the-art methods.
    \end{itemize}

    \begin{table}[tb]
      \centering\scriptsize
      \caption{Notation}
        \begin{tabular}{ll}     
        \toprule
        Notation                    & Explanation \\
        \midrule
        $\mathcal{U}$, $\mathcal{I}$, $\mathcal{I}^u$             & set of users, set of items, sequence of items of user $u$ \\
        $p$, $q$                    & positive item, negative item \\
        $\hat{x}_{upq}^t$           & preference of user $u$ towards item $p$ and $q$ at the $t$-th time step \\
        $d$                         & dimensionality of input vector \\
        $\boldsymbol{x}$, $\boldsymbol{h}$  & input and hidden state of GRU \\
        $\boldsymbol{x}_\mathrm{c}$, $\boldsymbol{h}_\mathrm{c}$              & contextual input, contextual hidden state \\
        $w_\mathrm{x}$, $w_\mathrm{h}$                & window widths \\
        $\boldsymbol{U}$, $\boldsymbol{W}$, $\boldsymbol{b}$    & transition matrices and bias of GRU \\
        $\boldsymbol{V}$                & transition matrix for $\boldsymbol{x}_c$ \\
        $\boldsymbol{E}$, $\boldsymbol{F}$  & embedding matrices for $\boldsymbol{h}$ and $\boldsymbol{h}_c$ \\
        $\boldsymbol{h}_\mathrm{o}$              & overall interest of $\boldsymbol{h}$ and $\boldsymbol{h}_c$ \\
        $\boldsymbol{r}_\mathrm{x}$, $\boldsymbol{Q}_\mathrm{x}$      & weight vector and matrix for input \\
        $\boldsymbol{r}_\mathrm{h}$, $\boldsymbol{Q}_\mathrm{h}$      & weight vector and matrix for hidden state \\
        $\boldsymbol{a}_\mathrm{x}$, $\boldsymbol{a}_\mathrm{h}$      & attention weights \\
        \bottomrule
        \end{tabular}
      \label{table:notation}
    \end{table}

\section{2 Related Work}
    \noindent
    In this section, we briefly review related works including sequential recommendation, contextual modeling and attention mechanism.

    The previous sequential methods mainly focus on Markov chains \cite{rendle2010factorizing}. Recently, recurrent neural network (RNN) based methods have become more powerful than the traditional sequential methods.
    Pooling-based representations of baskets are fed into RNN to make next basket recommendation \cite{yu2016dynamic}. Combined with the multiple external information, RNN can predict a user's next behavior more accurately \cite{liu2016context}. Incorporated with geographical distance and time interval information, RNN gains the state-of-the-art performance for location prediction \cite{liu2016predicting}. The RLBL model combines RNN and Log-BiLinear \cite{mnih2007three} to make multi-behavioral prediction \cite{liu2017multi}.
    Furthermore, extensions of RNN like long-short term memory (LSTM) \cite{hochreiter1997long} and gated recurrent unit (GRU) \cite{cho2014learning} are greatly developed, as they can better hold the long-term dependency \cite{Bengio1994Learning}. Here, our proposed network relies on GRU.

    Contextual information has been proven to be very important on different tasks.
    When learning word vectors in NLP, the prediction of next word is based on the sum or concatenation of recent input words \cite{le2014distributed}. About 3D action recognition, GCA-LSTM network applies all the hidden states in ST-LSTM \cite{liu2016spatio} to initialize global context memory by average pooling \cite{liu2017global}.
    On scene labeling, the patch is widely used. Episodic CAMN model makes contextualized representation for every referenced patch by using surrounding patches \cite{abdulnabiepisodic}.
    As the winner of the ImageNet object detection challenge of 2016, GBD-Net employs different resolutions and contextual regions to help identify objects \cite{zeng2016crafting}.
    Our contextual modeling is mainly focused on the input and hidden state of GRU, and we do not use any external information (e.g., time, location and weather) which is also called contextual information in recommender systems \cite{wang2012context,hariri2012context,liu2016context}.

    \begin{figure*}[tb]
    \centering
    \setlength{\abovecaptionskip}{-2pt}
    \setlength{\belowcaptionskip}{-7pt}
    \includegraphics[width=1\linewidth]{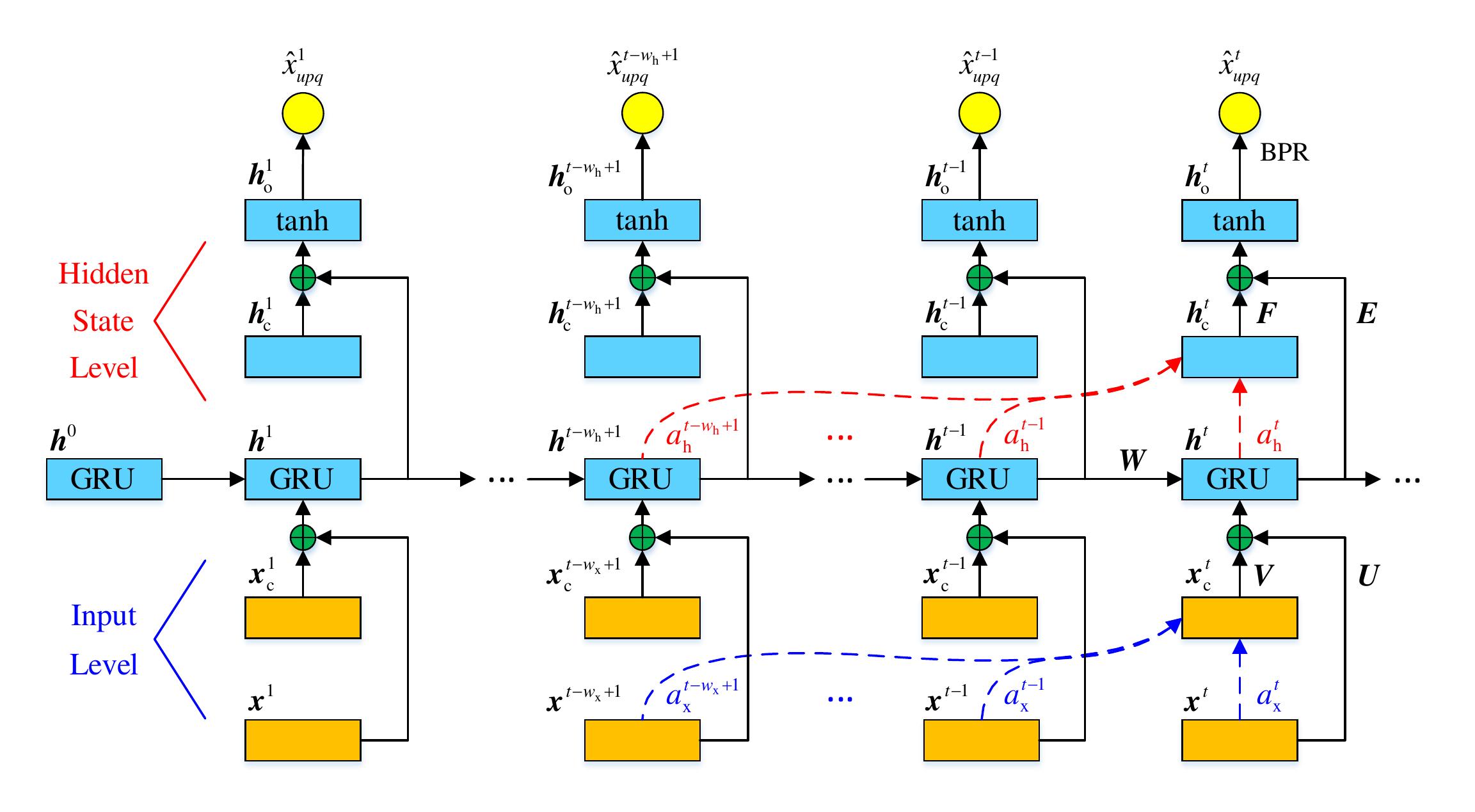}
    \caption{Diagram of HCA-GRU network (best viewed in color).
    Context of recent $w_x$ inputs is used to strengthen the current hidden state.
    Context of recent $w_h$ hidden states is used to construct the user's overall interest.
    We illustrate the whole modeling process at the $t$-th time step in detail and select the same window widths.}
    \label{fig:hca_gru}
    \end{figure*}

    Massive works benefit from the attention mechanism in recent years. Inspired by human perception, attention mechanism enables models to selectively concentrate on critical parts and construct a weighted sum. The mean pooling and max pooling, on the other hand, are unable to allocate appropriate credit assignment for individuals \cite{zhai2016deepintent}.
    This mechanism is first introduced in image classification to directly explore the space around the handwritten digits in the MNIST database\footnote{http://yann.lecun.com/exdb/mnist/} using visual attention \cite{mnih2014recurrent}.
    Besides using contextual modeling, Episodic CAMN applies the attention mechanism to adaptively choose important patches \cite{abdulnabiepisodic}.
    In NLP related tasks, attention mechanism is first introduced to do Neural Machine Translation (NMT) \cite{bahdanau2014neural}.
    This work jointly learns to align and translate words and conjectures the fixed-length vector (e.g., the last hidden state of source language) is a bottleneck for the basic encoder-decoder structure \cite{cho2014learning}.
    Another interesting work of NMT proposes a global attention model and a local attention model \cite{luong2015effective}. Besides, attention-based LSTM is proposed to automatically attend different parts of a sentence for sentiment classification \cite{wang2016attention}.

\section{3 Proposed HCA-GRU Network}
    \noindent
    In this section, we propose a novel network called Hierarchical Contextual Attention-based GRU (HCA-GRU). We first formulate the problem and introduce the basic GRU model. Then we present the contextual attention-based modeling on the input and hidden state respectively. Finally, we train the network with the BPR framework and the BPTT algorithm.

    \subsection{Problem Formulation}
    \noindent
    In order to simplify the problem formulation of sequential recommendation, we take buying histories of online shopping as an example. Use $\mathcal{U}=\{u_1,...,u_{|\mathcal{U}|}\}$ and $\mathcal{I}=\{i_1,...,i_{|\mathcal{I}|}\}$ to represent the sets of users and items respectively. Let $\mathcal{I}^u=(i^u_1,...,i^u_{|\mathcal{I}^u|})$ denote the items that the user $u$ has purchased in the time order.
    Given each user's history $I^u$, our goal is to predict a list of items that a user may buy in the near future. The notation is listed in Table \ref{table:notation} for clarity.

    \subsection{Gated Recurrent Unit}
    \noindent
    GRU is very effective to deal with the gradient vanishing and exploding problem. It has an $update$ gate $\boldsymbol{z}^t$ and a $reset$ gate $\boldsymbol{r}^t$ to control the flow of information. The formulas are:
    \begin{equation}    \label{eq_basic_gru}
    \begin{split}
    &\boldsymbol{z}^t = \sigma \left( \boldsymbol{U}_\mathrm{z} \boldsymbol{x}^t + \boldsymbol{W}_\mathrm{z} \boldsymbol{h}^{t-1} + \boldsymbol{b}_\mathrm{z} \right) \\
    &\boldsymbol{r}^t = \sigma \left( \boldsymbol{U}_\mathrm{r} \boldsymbol{x}^t + \boldsymbol{W}_\mathrm{r} \boldsymbol{h}^{t-1} + \boldsymbol{b}_\mathrm{r} \right) \\
    &\tilde{\boldsymbol{h}}^t = \tanh \left( \boldsymbol{U}_\mathrm{c} \boldsymbol{x}^t + \boldsymbol{W}_\mathrm{c} \left( \boldsymbol{r}^t \odot \boldsymbol{h}^{t-1} \right) + \boldsymbol{b}_\mathrm{c} \right) \\
    &\boldsymbol{h}^t = \left( 1 - \boldsymbol{z}^t \right) \odot \boldsymbol{h}^{t-1} + \boldsymbol{z}^t \odot \tilde{\boldsymbol{h}}^t
    \end{split}
    \end{equation}
    where $\boldsymbol{x}^t \in \mathbf{R}^d$ is the input vector and $t$ is the time step, the $logistic$ function $\sigma(x)=1/(1+e^{-x})$ is used to do non-linear projection, $\odot$ is the element-wise product between two vectors. $\tilde{\boldsymbol{h}}^t$ is the candidate state activated by element-wise $\tanh(x)$. The output $\boldsymbol{h}^t$ is the current hidden state.
    The last hidden state $\boldsymbol{h}^n$ is usually regarded as the sequence's representation, where $n$ is the length of the sentence.

    \subsection{HCA-GRU}
    \subsubsection{Contextual Attention-based Modeling on the Input}
    \noindent
    The first level of our HCA-GRU network is to model the complicated correlations among recent items and importantly improve the capability of user's short-term interest in each hidden state.
    This modeling process is represented in the bottom part of Figure. \ref{fig:hca_gru}.
    Please note that the subscripts $\mathrm{x}$ and $\mathrm{h}$ represent variables which are associated with the input and hidden state respectively.

    Take the current $t$-th time step for instance. Let $\boldsymbol{C}_\mathrm{x}^t$ be a context matrix consisting of recent $w_\mathrm{x}$ inputs, where $w_\mathrm{x}$ is the window width of the context. Then the following attention mechanism will generate a vector $\boldsymbol{a}_\mathrm{x}$ consisting of $w_\mathrm{x}$ weights and a weighted sum $\boldsymbol{x}_\mathrm{c}^t$ automatically reserving important information in $\boldsymbol{C}_\mathrm{x}^t$.
    \begin{subequations}    \label{eq_context_x}
    \begin{align}
    \boldsymbol{C}_\mathrm{x}^t &= \left[ \boldsymbol{x}^{t-w_\mathrm{x}+1}; \cdot \cdot \cdot ;\boldsymbol{x}^t \right]
        & \boldsymbol{C}_\mathrm{x}^t &\in \mathbf{R}^{w_\mathrm{x} \times d}  \label{eq_context_x_sub_context} \\
    \boldsymbol{e}_\mathrm{x}   &= \boldsymbol{r}_\mathrm{x}^\mathrm{T} \tanh \left( \boldsymbol{Q}_\mathrm{x} (\boldsymbol{C}_\mathrm{x}^t)^\mathrm{T} \right)
        & \boldsymbol{e}_\mathrm{x} &\in \mathbf{R}^{w_\mathrm{x}} \label{eq_context_x_sub_weights_origine} \\
    \boldsymbol{a}_\mathrm{x}   &= \mathrm{softmax} \left( \boldsymbol{e}_\mathrm{x} \right)
        & \boldsymbol{a}_\mathrm{x} &\in \mathbf{R}^{w_\mathrm{x}} \label{eq_context_x_sub_weights} \\
    \boldsymbol{x}_\mathrm{c}^t &= \left( \boldsymbol{a}_\mathrm{x} \boldsymbol{C}_\mathrm{x}^t \right)^\mathrm{T}
        & \boldsymbol{x}_c^t &\in \mathbf{R}^d \label{eq_context_x_sub_contextual_input}
    \end{align}
    \end{subequations}
    where $d$ is the size of the input vector, $\boldsymbol{r}_\mathrm{x} \in \mathbf{R}^d$  and $\boldsymbol{Q}_\mathrm{x} \in \mathbf{R}^{d \times d}$ are the weight vector and matrix for attention.
    $\boldsymbol{C}_\mathrm{x}^t$ is updated iteratively through the recurrent structure of GRU. The beginning of a sequence does not contain enough inputs and thus we apply zero vectors to make up this context matrix.
    $\boldsymbol{x}_\mathrm{c}^t$ is the contextual attention-based input.


    Next, we add $\boldsymbol{x}_c^t$ to the basic GRU in Eq. \ref{eq_basic_gru} to rewrite the formula as:
    \begin{equation}    \label{eq_context_x_hidden_layer}
    \begin{split}
    &\boldsymbol{z}^t = \sigma \left( \boldsymbol{U}_\mathrm{z} \boldsymbol{x}^t + \boldsymbol{V}_\mathrm{z} \boldsymbol{x}_c^t + \boldsymbol{W}_\mathrm{z} \boldsymbol{h}^{t-1} + \boldsymbol{b}_\mathrm{z} \right) \\
    &\boldsymbol{r}^t = \sigma \left( \boldsymbol{U}_\mathrm{r} \boldsymbol{x}^t + \boldsymbol{V}_\mathrm{r} \boldsymbol{x}_c^t + \boldsymbol{W}_\mathrm{r} \boldsymbol{h}^{t-1} + \boldsymbol{b}_\mathrm{r} \right) \\
    &\tilde{\boldsymbol{h}}^t = \tanh \left( \boldsymbol{U}_\mathrm{c} \boldsymbol{x}^t + \boldsymbol{V}_\mathrm{c} \boldsymbol{x}_\mathrm{c}^t + \boldsymbol{W}_c \left( \boldsymbol{r}^t \odot \boldsymbol{h}^{t-1} \right) + \boldsymbol{b}_\mathrm{c} \right) \\
    &\boldsymbol{h}^t = \left( 1 - \boldsymbol{z}^t \right) \odot \boldsymbol{h}^{t-1} + \boldsymbol{z}^t \odot \tilde{\boldsymbol{h}}^t
    \end{split}
    \end{equation}
    where $\boldsymbol{V} \in \mathbf{R}^{3 \times d \times d}$ formed by $\boldsymbol{V}_\mathrm{z}$, $\boldsymbol{V}_\mathrm{r}$ and $\boldsymbol{V}_\mathrm{c}$ is the transition matrix for $\boldsymbol{x}_c^t$. In this way, $\boldsymbol{h}^t$ contains not only information of original input $\boldsymbol{x}^t$ but also critical information of several recent inputs represented by $\boldsymbol{x}_\mathrm{c}^t$. The short-term interest in each hidden state is greatly enhanced.

    The $\boldsymbol{x}^t$ is still fed into the GRU unit in Eq. \ref{eq_context_x_hidden_layer} to guarantee the modeling of original long-term dependency in $\boldsymbol{h}^t$.
    However, $\boldsymbol{x}^t$ and $\boldsymbol{x}_\mathrm{c}^t$ are entered into Eq. \ref{eq_context_x_hidden_layer} together, which results in that $\boldsymbol{x}^t$ still has the highest weight among the multiple input items in $\boldsymbol{C}_\mathrm{x}^t$. We can see that this structure still follows the monotonic assumption of RNN, which will impede the modeling of complicated correlations among items.
    Hence, we execute further contextual attention-based modeling.

    \subsubsection{Contextual Attention-based Modeling on Hidden State}
    \noindent
    The second level of HCA-GRU is to further relieve the monotonic assumption problem and combine the long-term dependency and user's short-term interest to obtain current user's overall interest. The modeling process is illustrated in the upper part of Figure. \ref{fig:hca_gru}.

    The construction of contextual attention-based hidden state $\boldsymbol{h}_\mathrm{c}^t$ is similar to that of $\boldsymbol{x}_\mathrm{c}^t$:
    \begin{subequations}    \label{eq_context_h}
    \begin{align}
    \boldsymbol{C}_\mathrm{h}^t &= \left[ \boldsymbol{h}^{t-w_\mathrm{h}+1}; \cdot \cdot \cdot ;\boldsymbol{h}^t \right]
        & \boldsymbol{C}_\mathrm{h}^t &\in \mathbf{R}^{w_\mathrm{h} \times d}  \label{eq_context_h_sub_context} \\
    \boldsymbol{e}_\mathrm{h}   &= \boldsymbol{r}_h^\mathrm{T} \tanh \left( \boldsymbol{Q}_\mathrm{h} (\boldsymbol{C}_\mathrm{h}^t)^\mathrm{T} \right)
        & \boldsymbol{e}_\mathrm{h} &\in \mathbf{R}^{w_\mathrm{h}} \label{eq_context_h_sub_weights_origine} \\
    \boldsymbol{a}_\mathrm{h}   &= \mathrm{softmax} \left( \boldsymbol{e}_\mathrm{h} \right)
        & \boldsymbol{a}_\mathrm{h} &\in \mathbf{R}^{w_\mathrm{h}} \label{eq_context_h_sub_weights} \\
    \boldsymbol{h}_\mathrm{c}^t &= \left( \boldsymbol{a}_\mathrm{h} \boldsymbol{C}_\mathrm{h}^t \right)^T
        & \boldsymbol{h}_\mathrm{c}^t &\in \mathbf{R}^d \label{eq_context_h_sub_contextual_hidden_representation}
    \end{align}
    \end{subequations}
    where $\boldsymbol{C}_\mathrm{h}^t$ is the context matrix of recent $w_h$ hidden states. The operations on $\boldsymbol{C}_\mathrm{h}^t$ are the same with that on $\boldsymbol{C}_\mathrm{x}^t$. We consider $\boldsymbol{h}_\mathrm{c}^t$ as the current user's short-term interest within $w_\mathrm{h}$ time steps.

    The final representation is a non-linear combination of $\boldsymbol{h}_\mathrm{c}^t$ and $\boldsymbol{h}^t$. This is inspired by some works in NLP tasks, like recognizing textual entailment \cite{rocktaschel2015reasoning} and aspect-level sentiment classification \cite{wang2016attention}.
    \begin{equation}    \label{eq_context_h_overall_hidden_representation}
    \boldsymbol{h}_\mathrm{o}^t = \tanh \left( \boldsymbol{E}\boldsymbol{h}^t + \boldsymbol{F}\boldsymbol{h}_\mathrm{c}^t \right)
        \qquad \boldsymbol{h}_o^t \in \mathbf{R}^d
    \end{equation}
    where $\boldsymbol{E} \in \mathbf{R}^{d \times d}$ and $\boldsymbol{F} \in \mathbf{R}^{d \times d}$ are embedding matrices. $\boldsymbol{h}_\mathrm{o}^t$ is the current overall interest of the user.

%

    \begin{table*}[htbp]
      \centering\scriptsize
      \caption{Evaluation of different methods on two datasets with dimensionality $d=20$. We generate Top-5, 10, 15 and 20 items for each user.}
        \begin{tabular}{ccrrrrrrrrrrrr|c}

        \toprule    
        \multirow{2}*{dataset}  & \multirow{2}*{method}  &\multicolumn{3}{c}{@5}  &\multicolumn{3}{c}{@10} &\multicolumn{3}{c}{@15} &\multicolumn{3}{c}{@20} & \multirow{2}*{AUC}\\
        \cmidrule(lr){3-5}     \cmidrule(lr){6-8}   \cmidrule(lr){9-11} \cmidrule(lr){12-14}
          &   & Recall & MAP & NDCG & Recall & MAP & NDCG  & Recall & MAP & NDCG & Recall & MAP & NDCG \\

        \midrule
        \multirow{7}*{Taobao}
          & Random  & 0.0017 & 0.0007 & 0.0025 & 0.0029 & 0.0009 & 0.0028 & 0.0040 & 0.0009 & 0.0033 & 0.0057 & 0.0011 & 0.0042 & 50.000 \\
          & POP     & 0.0247 & 0.0176 & 0.0463 & 0.0999 & 0.0282 & 0.0845 & 0.1378 & 0.0320 & 0.1044 & 0.1676 & 0.0346 & 0.1190 & 58.612 \\
          & BPR     & 0.3728 & 0.4756 & 0.8604 & 0.5597 & 0.5013 & 0.8717 & 0.7230 & 0.5128 & 0.9247 & 0.8812 & 0.5194 & 0.9769 & 65.718 \\
          & GRU     & 0.4777 & 0.5850 & 1.0878 & 0.8101 & 0.6149 & 1.0990 & 1.1233 & 0.6273 & 1.1783 & 1.3977 & 0.6343 & 1.2718 & \textbf{66.218} \\
          & \textbf{HCA-GRU-x3} & 0.7236 & 0.7360 & 1.5484 & 1.1491 & 0.8124 & 1.6099 & 1.4796 & 0.8518 & 1.7387 & 1.7764 & 0.8799 & 1.8624 & 65.796 \\
          & \textbf{HCA-GRU-h5} & 0.7347 & 0.7459 & 1.5686 & 1.1452 & 0.8205 & 1.6001 & 1.4723 & 0.8634 & 1.7268 & 1.7605 & 0.8920 & 1.8452 & 65.927 \\
          & \textbf{HCA-GRU-x5-h5} & \textbf{0.7650} & \textbf{0.7570} & \textbf{1.6176} & \textbf{1.1909} & \textbf{0.8377} & \textbf{1.6588} & \textbf{1.5197} & \textbf{0.8788} & \textbf{1.7869} & \textbf{1.8235} & \textbf{0.9067} & \textbf{1.9131} & 65.774 \\

        \midrule
        \multirow{7}*{Outbrain}
          & Random  & 0.007  & 0.003  & 0.006  & 0.015  & 0.004  & 0.009  & 0.022  & 0.005  & 0.012  & 0.031  & 0.005  & 0.014  & 50.000 \\
          & POP     & 0.026  & 0.012  & 0.019  & 1.188  & 0.182  & 0.535  & 1.191  & 0.183  & 0.536  & 1.511  & 0.209  & 0.648  & 83.291\\
          & BPR     & 4.092  & 2.637  & 4.050  & 8.523  & 3.380  & 5.992  & 23.750 & 5.215  & 10.935 & \textbf{29.209} & 5.740  & 12.555 & 86.982 \\
          & GRU     & 11.145 & 6.118  & 9.451  & 17.560 & 7.319  & 11.987 & 21.413 & 7.551  & 12.969 & 23.492 & 8.012  & 14.257  & 91.546 \\
          & \textbf{HCA-GRU-x2} & 13.485 & 8.862 & 13.237 & 19.033 & 9.970 & 15.551 & 21.744 & 10.277 & 16.498 & 24.077 & 10.502 & 17.316 & 91.539 \\
          & \textbf{HCA-GRU-h3} & 14.562 & 8.820 & 12.988 & 19.993 & 9.979 & 15.353 & 24.010 & 10.381 & 16.522 & 26.233 & 10.652 & 17.361 & 91.428 \\
          & \textbf{HCA-GRU-x2-h3} & \textbf{16.396} & \textbf{10.136} & \textbf{14.225} & \textbf{23.328} & \textbf{11.520} & \textbf{17.097} & \textbf{26.836} & \textbf{11.904} & \textbf{18.280} & 29.188 & \textbf{12.123} & \textbf{19.055} & \textbf{91.565} \\

        \bottomrule
        \end{tabular}
      \label{table:result}
    \end{table*}

    \subsubsection{Network Learning}
    \noindent
    The proposed network can be trained under the BPR framework and by using the classical BPTT algorithm.
    BPR is a powerful pairwise method for implicit feedback. Many RNN based methods have successfully applied BPR to train their models \cite{liu2016predicting,yu2016dynamic,liu2017multi}.

    The training set $S$ is formed by $(u,p,q)$ triples:
    \begin{equation}    \label{dataset}    
    S = \left\{ {( u,p,q )} | {u \in \mathcal{U}} \wedge {p \in \mathcal{I}^u} \wedge {q \in \mathcal{I}\setminus \mathcal{I}^u} \right\}
    \end{equation}
    where $u$ denotes the user, $p$ and $q$ represent the positive and negative items respectively. Item $p$ is from the user's history $\mathcal{I}^u$, while item $q$ is randomly chosen from the rest items.
    Then we calculate the user's preference for positive and negative items based on the current user's overall interest. At the $t$-th time step, the preference is:
    \begin{equation}   \label{eq_preference}  
    \hat{x}_{upq}^t =
            {\left( \boldsymbol{h}_\mathrm{o}^t \right)}^\mathrm{T}
            \left( \boldsymbol{x}_p^{t+1} - \boldsymbol{x}_q^{t+1} \right)
    \end{equation}
    where $\boldsymbol{x}_p^{t+1}$ and $\boldsymbol{i}_q^{t+1}$ represent next positive and negative inputs respectively.

    The objective function minimizes the following formula:
    \begin{equation}    \label{eq_optimization_bpr}    
    \Theta^* = \underaccent{\text{$\Theta$}}{\text{argmin}} \sum_{( u,p,q ) \in S}-\ln{\left( \sigma \left(
    \hat{x}_{upq} \right) \right)} + \frac{\lambda_\Theta}{2}{\| \Theta \|^2}
    \end{equation}
    where $\Theta=\{\boldsymbol{X}, \boldsymbol{U}, \boldsymbol{W}, \boldsymbol{V}, \boldsymbol{b}, \boldsymbol{r}_\mathrm{x}, \boldsymbol{Q}_\mathrm{x},  \boldsymbol{r}_\mathrm{h}, \boldsymbol{Q}_\mathrm{h}, \boldsymbol{E}, \boldsymbol{F} \}$ is the set of parameters. $\boldsymbol{X}$ is all latent features of items. $\boldsymbol{U}$ is a matrix formed by $\boldsymbol{U}_\mathrm{z}$, $\boldsymbol{U}_\mathrm{r}$ and $\boldsymbol{U}_\mathrm{c}$ used in Eqs. \ref{eq_basic_gru} and \ref{eq_context_x_hidden_layer}, which is similar to $\boldsymbol{W}$, $\boldsymbol{V}$ and $\boldsymbol{b}$.
    $\lambda_\Theta \geqslant 0$ is the regularization parameter.
    Then, HCA-GRU can be learned by the stochastic gradient descent and BPTT. The parameters are automatically updated by Theano \cite{bergstra2010theano}. We consider a whole user sequence as a mini-batch.

    During the test process, we need to recompute each user's final overall interest $\boldsymbol{h}_\mathrm{o}^n$ by using the fixed parameter $\Theta$. We first redo contextual attention-based modeling on the input at each time step, and the modeling on hidden state only needs to be made at the last time step on $\boldsymbol{h}^n$ to obtain $\boldsymbol{h}_c^n$. Then we acquire $\boldsymbol{h}_\mathrm{o}^n = \tanh \left( \boldsymbol{E}\boldsymbol{h}^n + \boldsymbol{F}\boldsymbol{h}_\mathrm{c}^n \right)$.

\section{4 Experimental Results and Analysis}
    \noindent
    In this section, we conduct experiments on two real-world datasets. First, we introduce the datasets, evaluation metrics and baseline methods. Then we make comparison between HCA-GRU and the baseline methods. Finally, we present the window width selection and attention visualization.

    \subsection{Experimental Settings}
    \noindent
    \subsubsection{Datasets}
    \noindent
    Experiments are carried out on two datasets collected from \emph{Taobao}\footnote{https://tianchi.shuju.aliyun.com/datalab/dataSet.htm?id=13} and \emph{Outbrain}\footnote{https://www.kaggle.com/c/outbrain-click-prediction}.
    \begin{itemize}
        \item
            \textbf{Taobao} is a dataset for clothing matching competition on \emph{TianChi}\footnote{https://tianchi.aliyun.com/} platform. User historical behavior data is applied to make sequential recommendation. We hold users who purchase at least 30 times ($|\mathcal{I}^u|\geqslant$30). Furthermore, similar to the p-RNNs model \cite{hidasi2016parallel}, we filter out the very long sequences, because the users with too long sequences may scalp clothing products.
        \item
            \textbf{Outbrain} is a dataset for click prediction on Kaggle. This competition asks contestants to predict which recommended content on pages that each user will click.
            We only apply a sample version of the tremendous page views log data and we choose users who have more than 10 views ($|\mathcal{I}^u|\geqslant$10).
    \end{itemize}
    The basic statistics of two datasets are listed in Table \ref{table:datasets}. Both datasets have massive sequential implicit feedbacks.
    Intuitively, Taobao has a much larger number of items than Outbrain, which will naturally result in a huge search space when making recommendation.  and will potentially degrade the performance.

    \begin{table}[tb]
      \centering\scriptsize
      \caption{Datasets. We list the numbers of users, items, feedbacks, average sequence length and sparsity of both dataset.}
        \begin{tabular}{crrrcc}     
        \toprule
        Dataset     & \#users & \#items & \#feedbacks & \#avg. seq. len.  & sparsity (\%) \\
        \midrule
        Taobao      & 36,986  & 267,948   & 1,640,433  & 44.35  & 99.9834 \\
        Outbrain    & 65,573  & 69,210   & 833,646    & 12.71  & 99.9816 \\
        \bottomrule
        \end{tabular}%
      \label{table:datasets}%
    \end{table}%

    \subsubsection{Evaluation Metrics}
    Performance is evaluated on the test set under metrics consisting of Recall, Mean Average Precision (MAP) \cite{manning2008introduction} and Normalized Discounted Cumulative Gain (NDCG) \cite{wang2015learning}. The former one is an evaluation of unranked retrieval sets while the latter two are evaluations of ranked retrieval results. Here we consider Top-$k$ (e.g., $k~=~$5, 10, 15, 20) recommendation. The top-bias property of MAP and NDCG is significant for recommendation \cite{shi2012tfmap}.
    Besides, the Area Under the ROC Curve (AUC) \cite{he2016vbpr} is introduced to evaluate the global performance.
    We select the first 80\% of each user sequence as the training set and the rest 20\% as the test set.
    Besides, we remove the duplicate items in each user's test sequence.
    \begin{figure*}[htbp]
    \centering
    \setlength{\abovecaptionskip}{0pt}
    \setlength{\belowcaptionskip}{0pt}
    \subfigure[Taobao]{
    \begin{minipage}[b]{1\textwidth}
    \includegraphics[width=1\textwidth]{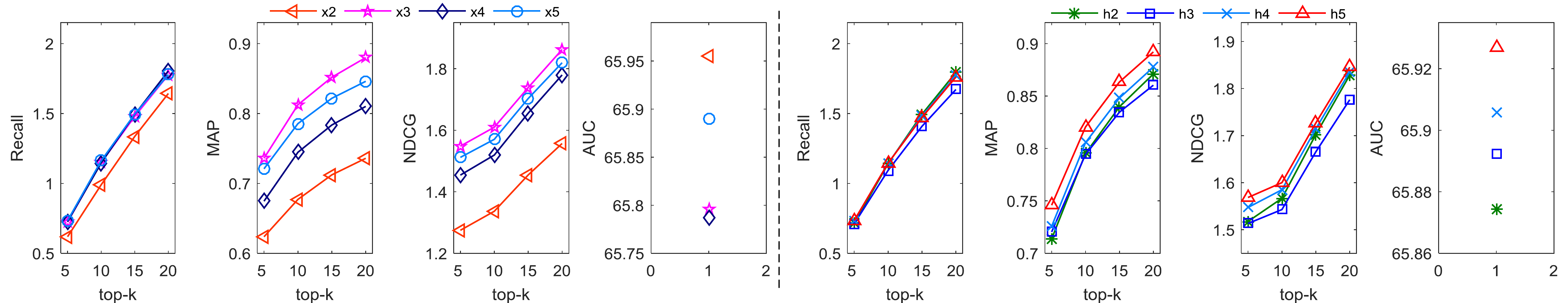}
    \label{figsub:taobao_Recall_varied_dimension_at_30}
    \end{minipage}
    }
    \subfigure[Outbrain]{
    \begin{minipage}[b]{1\textwidth}
    \includegraphics[width=1\textwidth]{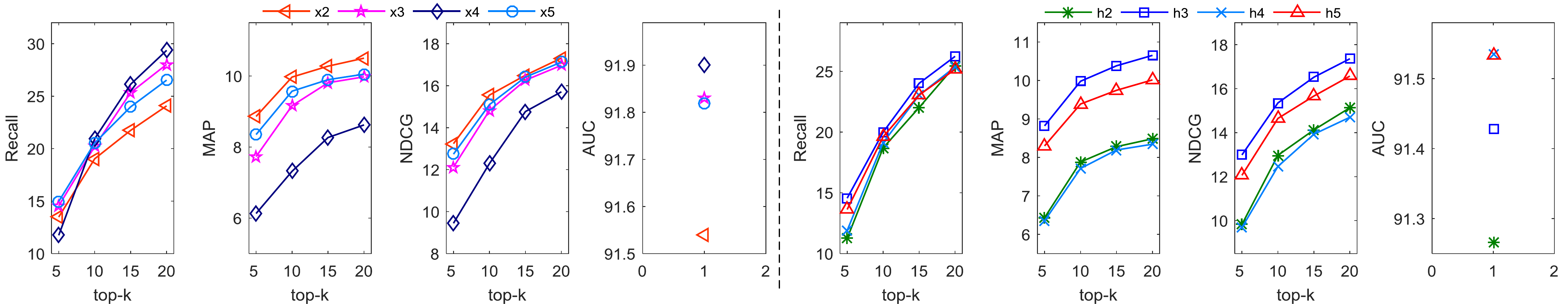}
    \label{figsub:amazon_Recall_varied_dimension_at_30}
    \end{minipage}
    }
    \caption{Evaluation of different window widths of input and hidden state on two datasets. We generate Top-5, 10, 15 and 20 items. Subnetworks are HCA-GRU-x2$\sim$5 and HCA-GRU-h2$\sim$5, and we omit the prefix `HCA-GRU-' in the figures.}
    \label{fig:window_x_or_window_h}
    \end{figure*}

    \subsubsection{Comparison}
    \noindent
    We compare HCA-GRU with several comparative methods:
    \begin{itemize}
        \item
            \textbf{Random}: Items are randomly selected for all users. The AUC of this baseline is 0.5 \cite{rendle2009bpr}.
        \item
            \textbf{POP}: This method recommends the most popular items in the training set to users.
        \item
            \textbf{BPR}: This method refers to the BPR-MF for implicit feedback \cite{rendle2009bpr}. This BPR framework is the state-of-the-art among pairwise methods and applied to many tasks.
        \item
            \textbf{GRU}: RNN is the state-of-the-art for sequential recommendation \cite{yu2016dynamic}. We apply an extension of RNN called GRU for capturing the long-term dependency.
    \end{itemize}


    Our proposed complete network is abbreviated as $HCA$-$GRU$-$\mathrm{x}w_\mathrm{x}$-$\mathrm{h}w_\mathrm{h}$, where $w_\mathrm{x}$ and $w_\mathrm{h}$ are values of window widths. When we only conduct contextual attention-based modeling on either the input or hidden state, the subnetwork is abbreviated as $HCA$-$GRU$-$\mathrm{x}w_\mathrm{x}$ or  $HCA$-$GRU$-$\mathrm{h}w_\mathrm{h}$.
    Besides, parameter $\Theta$ is initialized by a same uniform distribution $[-0.5, 0.5]$.
    The dimensionality is set as $d=20$.
    The learning rate is set as $\alpha=0.01$ for all methods. Regularizers are set as $\lambda_\Theta=0.001$ for all parameters.

    \subsection{Analysis of Experimental Results}
    \noindent
    Table \ref{table:result} illustrates performances of all compared methods on two datasets. We list the performance of HCA-GRU network under best window widths which are discussed in the next subsection.
    Please note that all values in Tables \ref{table:result}, \ref{table:window_x_and_window_h} and Figure \ref{fig:window_x_or_window_h} are represented in percentage, and the '\%' symbol is omitted in these tables and figures. All methods have run four times and the averages are computed as the performance values.

    From a global perspective, our HCA-GRU network is very effective and significantly outperforms the basic methods. Within the HCA-GRU, the subnetwork performances of contextual attention-based modeling on the input and hidden state are very close, and the latter is slightly better than the former. The complete HCA-GRU obtains the best performance.
    Then on the four metrics, HCA-GRU performs differently. It has a great improvement on Recall, but the improvement is visibly getting smaller from Top-$5$ to Top-$20$. In contrast, the improvements on MAP and NDCG are very significant and quite stable. In most cases, there are almost 50\% improvements.
    HCA-GRU has a very prominent performance of sorting. This is because HCA-GRU can effectively capture the complicated correlations among several adjacent items and distinguish which items the user are more interested in.
    Relatively unexpected, compared with GRU, HCA-GRU has no promotion on AUC. Incorporating external information may be a possible way to obtain the improvement.
    Next, HCA-GRU does very well under different datasets. The absolute values on two datasets are very different. This is because the huge items set of Taobao greatly hinders the search performance. But HCA-GRU obtains similar relative improvements on two datasets, which exactly exhibits the effectiveness of HCA-GRU.

    \begin{table*}[htbp]
      \centering\scriptsize
      \caption{Evaluation of different window widths for complete HCA-GRU network on two datasets.}
        \begin{tabular}{cccccccccccccc|c}

        \toprule
        \multirow{2}*{dataset}  & \multirow{2}*{method}  &\multicolumn{3}{c}{@5}  &\multicolumn{3}{c}{@10} &\multicolumn{3}{c}{@15} &\multicolumn{3}{c}{@20} & \multirow{2}*{AUC}\\
        \cmidrule(lr){3-5}     \cmidrule(lr){6-8}   \cmidrule(lr){9-11} \cmidrule(lr){12-14}
          &   & Recall & MAP & NDCG & Recall & MAP & NDCG  & Recall & MAP & NDCG & Recall & MAP & NDCG \\

        \midrule
        \multirow{8}*{Taobao}
          & HCA-GRU-x3-h2 & 0.6925 & 0.7060 & 1.4724 & 1.0841 & 0.7801 & 1.5206 & 1.3919 & 0.8199 & 1.6388 & 1.6590 & 0.8467 & 1.7504 & 65.797 \\
          & HCA-GRU-x3-h3 & 0.6751 & 0.7182 & 1.4759 & 1.0543 & 0.7874 & 1.5168 & 1.3716 & 0.8263 & 1.6460 & 1.6510 & 0.8529 & 1.7621 & 65.825 \\
          & HCA-GRU-x3-h4 & 0.6782 & 0.7172 & 1.4723 & 1.0511 & 0.7867 & 1.5148 & 1.3822 & 0.8262 & 1.6439 & 1.6662 & 0.8526 & 1.7632 & 65.737 \\
          \cmidrule(lr){2-2}
          & HCA-GRU-h5-x2 & 0.7146 & 0.7306 & 1.5263 & 1.1125 & 0.8062 & 1.5700 & 1.4529 & 0.8492 & 1.7011 & 1.7449 & 0.8779 & 1.8237 & 65.697 \\
          & HCA-GRU-h5-x3 & 0.7349 & 0.7368 & 1.5563 & 1.1698 & 0.8167 & 1.6132 & \textbf{1.5329} & 0.8606 & 1.7534 & \textbf{1.8290} & 0.8910 & 1.8736 & 65.774 \\
          & HCA-GRU-h5-x4 & 0.7349 & 0.7409 & 1.5673 & 1.1406 & 0.8142 & 1.6043 & 1.4823 & 0.8560 & 1.7400 & 1.7785 & 0.8825 & 1.8637 & 65.759 \\
          & \textbf{HCA-GRU-h5-x5} & \textbf{0.7650} & \textbf{0.7570} & \textbf{1.6176} & \textbf{1.1909} & \textbf{0.8377} & \textbf{1.6588} &1.5197 & \textbf{0.8788} & \textbf{1.7869} & 1.8235 & \textbf{0.9067} & \textbf{1.9131} & 65.844 \\

        \midrule
        \multirow{8}*{Outbrain}
          & HCA-GRU-x2-h2 & 13.515 & 8.902 & 12.902 & 20.706 & 10.245 & 15.800 & 24.194 & 10.788 & 17.173 & 27.069 & 11.064 & 18.029 & 91.355 \\
          & \textbf{HCA-GRU-x2-h3} & 16.396 & \textbf{10.136} & 14.225 & \textbf{23.328} & \textbf{11.520} & \textbf{17.097} & \textbf{26.836} & \textbf{11.904} & \textbf{18.280} & 29.188 & \textbf{12.123} & \textbf{19.055} & 91.565 \\
          & HCA-GRU-x2-h4 & \textbf{16.698} & 10.103 & \textbf{14.306} & 21.721 & 11.210 & 16.747 & 25.893 & 11.665 & 18.106 & 28.204 & 11.965 & 18.839 & 91.516 \\
          & HCA-GRU-x2-h5 & 14.799 & 7.998 & 11.886 & 22.144 & 9.450 & 14.992 & 26.750 & 10.105 & 16.519 & \textbf{29.719} & 10.356 & 17.414 & 91.652 \\
          \cmidrule(lr){2-2}
          & HCA-GRU-h3-x3 & 13.183 & 7.251 & 10.883 & 20.448 & 8.704 & 14.070 & 25.481 & 9.201 & 15.501 & 28.994 & 9.524 & 16.637 & 91.660 \\
          & HCA-GRU-h3-x4 & 12.482 & 6.250 &  9.733 & 21.996 & 7.873 & 13.512 & 25.712 & 8.457 & 14.974 & 28.645 & 8.696 & 15.895 & \textbf{91.704} \\
          & HCA-GRU-h3-x5 & 12.368 & 7.232 & 10.869 & 20.264 & 8.667 & 14.161 & 25.073 & 9.352 & 15.946 & 28.509 & 9.656 & 16.990 & 91.659 \\

        \bottomrule
        \end{tabular}
      \label{table:window_x_and_window_h}
    \end{table*}

    \begin{table*}[tb]
      \centering\scriptsize
      \caption{Examples of attention weights of two users' sequences on two datasets respectively.
      On Taobao, we apply the best HCA-GRU-x5-h5 network.
      During the test process, we only need to do contextual attention-based modeling on the last hidden state $h^n$, so we have 5 weights for the last 5 hidden states.
      The window width of contextual input is also 5, so the last 5 hidden states each have 5 weights of 5 inputs.
      On Outbrain, HCA-GRU-x2-h3 is used.}
        \begin{tabular}{c|ccccccccc|cccc}

        \toprule
          &\multicolumn{9}{c}{Taobao}  &\multicolumn{4}{c}{Outbrain} \\

        \midrule
          Hidden states
          & $h^{n-8}$ & $h^{n-6}$ & $h^{n-6}$ & $h^{n-5}$ & $h^{n-4}$ & $h^{n-3}$  & $h^{n-2}$ & $h^{n-1}$ & $h^{n}$ & $h^{n-3}$  & $h^{n-2}$ & $h^{n-1}$ & $h^{n}$\\
        \cmidrule(lr){2-14}
          Weights
          &  &  &  &  & \textbf{0.2070} & 0.1873 & 0.2026 & 0.1991 & 0.2039 &  & \textbf{0.3659} & 0.3270 & 0.3070 \\

        \midrule
          Inputs
          & $x^{n-8}$ & $x^{n-6}$ & $x^{n-6}$ & $x^{n-5}$ & $x^{n-4}$ & $x^{n-3}$  & $x^{n-2}$ & $x^{n-1}$ & $x^{n}$ & $x^{n-3}$  & $x^{n-2}$ & $x^{n-1}$ & $x^{n}$ \\
        \cmidrule(lr){2-14}
          \multirow{5}*{Weights}
          & 0.1703 & 0.2709 & \textbf{0.2894} & 0.1437 & 0.1257 &  &  &  &  & \textbf{0.6605} & 0.3395 &  &  \\
          &  & 0.3229 & \textbf{0.3450} & 0.1713 & 0.1499 & 0.0109 &  &  &  &  & \textbf{0.5974} & 0.4026 &  \\
          &  &  & \textbf{0.3955} & 0.1963 & 0.1718 & 0.0125 & 0.2238 &  &  &  &  & 0.3189 & \textbf{0.6811} \\
          &  &  &  & 0.2517 & 0.2202 & 0.0161 & \textbf{0.2870} & 0.2250 &  &  &  &  &  \\
          &  &  &  &  & 0.2898 & 0.0212 & \textbf{0.3776} & 0.2960 & 0.0154 &  &  &  &  \\

        \bottomrule
        \end{tabular}
      \label{table:attention_weights}
    \end{table*}

    \subsection{Analysis of Window Width}
    \noindent
    In this subsection, we investigate the best window width for HCA-GRU. Both window widths $w_\mathrm{x}$ and $w_\mathrm{h}$ range in $[2, 3, 4, 5]$. The basic GRU can be considered as a special case of our network when both $w_\mathrm{x}$ and $w_\mathrm{h}$ are 1.

    \subsubsection{Subnetwork} First we select best window widths for two subnetworks on two datasets. Figure \ref{fig:window_x_or_window_h} illustrates the performance of four metrics.
    For two subnetworks with different window widths, the difference is slight on Recall and AUC, while there is an obvious difference on MAP and NDCG. Consequently, best window widths for two subnetworks are chosen as $w_\mathrm{x}=3,~w_\mathrm{h}=5$ on Taobao and $w_\mathrm{x}=2,~w_\mathrm{h}=3$ on Outbrain respectively according to MAP and NDCG. Both $w_\mathrm{x}$ and $w_\mathrm{h}$ on Taobao are larger than those on Outbrain. It may be because the average sequence length on Taobao is longer.

    \subsubsection{Complete Network}
    Based on the best window widths of subnetworks, we pick proper $w_\mathrm{x}$ and $w_\mathrm{h}$ for complete HCA-GRU network. Grid search over combinations of $w_\mathrm{x}\in[2, 3, 4, 5]$ and $w_\mathrm{h}\in[2, 3, 4, 5]$ is very time-consuming. Instead, we fix one with its best value and adjust the other. The results are shown in Table \ref{table:window_x_and_window_h}.

    Generally, the combination of best window widths from the subnetworks can generate good results but does not guarantee the best. Combinations $w_\mathrm{x}=5,~w_\mathrm{h}=5$ and $w_\mathrm{x}=2,~w_\mathrm{h}=3$ are best for Taobao and Outbrain respectively.
    Then we look into each dataset. On Taobao, the overall performance of several HCA-GRU networks is better if we fix $w_\mathrm{h}=5$ and adjust $w_\mathrm{x}$ than if we fix $w_\mathrm{x}=3$ and adjust $w_\mathrm{h}$. On Outbrain, the situation is contrary.
    Therefore, we can not say that one of the two subnetworks must be better than the other.

    \subsection{Analysis of Attention Weights}
    \noindent
    The attention mechanism generates a vector to summarize the contextual information. The attention weights can be obtained in Eqs. \ref{eq_context_x_sub_weights} and \ref{eq_context_h_sub_weights}. We take the networks HCA-GRU-x5-h5 on Taobao and HCA-GRU-x2-h3 on Outbrain for instance. Weights of two users' sequences at the last $n$-th time step in the test process are listed in Table \ref{table:attention_weights}.

    On one hand, we focus on weights of inputs. First, we look at each line of weights. Obviously, attention can capture the most important item with the highest weight. Next, we check each column. If an item is very important in a context, it would be probably also very important in the next context, and vice versa. For example, item $x^{n-6}$ and $x^{n-2}$ on Taobao are very important, and the weights of $x^{n-3}$ are all less than 3\%.
    On the other hand, we study weights of hidden states. The weights have little difference with each other on Taobao and hold a monotonic descending order on Outbrain respectively.
    To sum up, our HCA-GRU breaks through the limitation of monotonic assumption of RNN. The attention mechanism can capture the most important information and assign time-independent nonmonotonic weights.


\section{5 Conclusion}
    \noindent
    In this work, we have proposed a novel network called hierarchical contextual attention-based GRU (HCA-GRU) for sequential recommendation.
    HCA-GRU can relieve the problem of monotonic temporal dependency of RNN. The hierarchical architecture is effective to obtain user's overall interest built by long-term dependency and short-term interest. The contextual attention-based modeling on the input and hidden state can capture the important information and acquire time-independent nonmonotonic weights. This greatly enhances the modeling of user's short-term interest.
    Experiments verify the state-of-the-art performance of our network, especially the performance of sorting.

\bibliography{aaai18References}   
\bibliographystyle{aaai}

\end{document}